\documentstyle[aps,prl,graphicx,floats,amssymb]{revtex}
\DeclareGraphicsExtensions{.eps,.ps}
\begin{document}
\draft

\wideabs{

  \title{Unusual thermodynamical and transport signatures of the BCS
    Bose-Einstein crossover scenario below $T_c$}

  \author{Qijin Chen, Ioan Kosztin, and K. Levin}

  \address{James Franck Institute, University of Chicago, 5640 South
    Ellis Avenue, Chicago, Illinois 60637, USA}

  \date{\today} 

\maketitle
%----------------------------------------------------------------
\begin{abstract}
  In this paper we present predictions for thermodynamic and transport
  properties of a BCS Bose-Einstein crossover theory, below $T_c$, which
  theory satisfies the reasonable constraints that it yield (i) the Leggett
  ground state and (ii) BCS theory at weak coupling and all temperatures
  $T$.  The nature of the strong coupling limit is inferred, along with the
  behavior of the Knight shift, superfluid density, and specific heat.
  Comparisons with existing data on short coherence length superconductors,
  such as organic and high $T_c$ systems, are presented,
  which provide some support for the present picture.
\end{abstract}

\pacs{PACS numbers: %
74.20.-z, % Theories and models of superconducting state
74.25.Bt, %Thermodynamic properties
74.25.Fy, %Transport properties
74.25.Nf%, %Response to electromagnetic fields (nuclear magnetic resonance, 
         %surface impedance, etc.)
\hfill \textsf{\textbf{cond-mat/9908362}}%
}
\vskip -3mm
}

%------------------------------------------------------
Within the pseudogap regime of the cuprates, it has been widely argued that
the excitations of the superconducting state are either fermionic\cite{Lee}
or bosonic\cite{Emery} in character. In this paper we discuss a third
scenario (associated with the BCS Bose-Einstein condensation (BEC) crossover
approach), in which the excitations contain a mix of bosonic and fermionic
properties\cite{Chen}.

The BCS-BEC crossover scheme has been viewed as relevant to the cuprates and
other ``exotic"\cite{Uemura,Randeria} superconductors where their short
coherence lengths, $\xi$, naturally lead to a breakdown of strict BCS
theory.  The BCS-BEC scenario \cite{NSR,Haussmann,Micnas,Tschern,Chen} owes
its origin to Eagles\cite{Eagles} and Leggett\cite{Leggett} who proposed a
ground state wavefunction of the BCS form $\Psi_0 = \Pi_{\mathbf{k}} (
u_{\mathbf{k}} + v_{\mathbf{k}} c^\dagger_{\mathbf{k}\uparrow}
c^\dagger_{\mathbf{-k}\downarrow} )|0\rangle $, which describes the
continuous evolution between a BCS system, having weak coupling $g$ and
large $\xi$, towards a BEC system with large $g$ and small $\xi$.  Here $
u_{\mathbf{k}}, v_{\mathbf{k}}$ are the standard coherence factors of BCS
theory, which are determined \textit{in conjunction} with
the number constraint.

The essence of this paper is a characterization of the excitations of
$\Psi_0$ and their experimental signatures (for all $T \le T_c$).  New
thermodynamical effects stemming from bosonic degrees of freedom must
necessarily enter, as one crosses out of the BCS regime, towards BEC.  Here
we show that the bosonic excitations (which appear in the gap equations and
which we call ``pairons") are different from the collective phase
mode\cite{Kosztin2}: they generally have a quadratic dispersion similar to
that of a \textit{quasi-ideal} BEC system --- a consequence of the mean
field treatment of the pairs which, in turn, is dictated by the general mean
field character of $\Psi_0$.  These ideas are applied to the cuprates and
other short $\xi$ superconductors.

We consider fermions, with lattice dispersion $\epsilon_{\mathbf{k}}$,
(measured with respect to the fermionic chemical potential $\mu$), and
with interaction $V_{\mathbf{k,k'}} = g\varphi_{\mathbf{k}}
\varphi_{\mathbf{k'}}$, where $g<0$; here $\varphi_{\mathbf{k}}=1$ and
$(\cos k_x -\cos k_y)$ for $s$- and $d$-wave pairing, respectively.  We
begin by reviewing BCS theory, in terms of a somewhat unfamiliar but
very useful formalism\cite{Kadanoff} which is then
generalized\cite{Chen,Kosztin} to the crossover problem.
For brevity, we use a four-momentum notation: $K=({\bf k}, i\omega),
\sum_K = T \sum _ {{\bf k}, \omega}$, etc. We also suppress
$\varphi_{\bf k}$ until the final equations.

BCS theory involves the pair susceptibility $ \chi (Q) = \sum_K G (K)
G_0 (Q-K) $, where the Green's function $G$ satisfies $G^{-1} = G_0
^{-1} + \Sigma $, with order parameter $\Delta_{sc}$ and $ \Sigma(K) =
-\Delta_{sc}^2 G_0(-K)$.  In this notation, the gap equation is
\begin{equation} 
1 + g \chi (0) = 0,\qquad T \le T_c.
  \label{eq:gap0} 
\end{equation}  
At $Q = 0$, the summand in $\chi$ is the Gor'kov ``$F$" function (up to
a multiplicative factor $\Delta_{sc}$) and \textit{this serves to
  highlight the central role played in BCS theory by the more general
  quantity} $G(K)G_0(Q-K)$. Note that (for $ Q \ne 0$), $\chi (Q)$ is
\textit{distinct from the pair susceptibility of the collective phase
  mode} which enters as $ \sum_K \{ G (K )[ G (Q-K) + G ( -Q -K)] +2F
(K) F (K-Q)\} $ \cite{Kosztin2,Kulik}. Here each Gor'kov ``$F$''
function introduces one $GG_0$, so that \textit{the collective mode
  propagator depends on effectively higher order Green's functions than
  does the gap equation}.

The observations in italics were first made in
Ref.~\onlinecite{Kadanoff} where it was noted that the BCS gap equation
could be rederived by truncating the equations of motion so that only
the one ($G$) and two particle (${\cal T}$) propagators appeared.  Here
$G$ depends on $\Sigma$ which in turn depends on $ {\cal T} $. In
general ${\cal T}$ has two additive contributions\cite{Kosztin}, from
the condensate (sc) and the non-condensed (pg) pairs. Similarly the
associated self energy\cite{Kadanoff} $\Sigma (K) = \sum_Q {\cal{T}} (Q)
G_0(Q-K)$ can be decomposed into $\Sigma_{pg}(K) + \Sigma_{sc}(K)$.  The
two contributions in $\Sigma$ come respectively from ${\cal{T}}_{sc}(Q)
= -\Delta_{sc}^2 \delta (Q)/T$, and from the $Q \ne 0$ pairs, with
${\cal{T}}_{pg}(Q) = g/(1 + g \chi (Q))$.  In the leading order mean
field theory $\Sigma = \Sigma_{sc}$ $= -\Delta_{sc} ^2 G_0 (-K)$ which,
from Eq.~(1), yields the usual BCS gap equation.

More generally, at larger $g$, the above equations hold but we now
include feedback into Eq.~(1) from the finite momentum pairs, via
$\Sigma_{pg} (K) = \sum_Q {\cal{T}}_{pg}(Q) G_0 (Q- K) \approx G_0 (-K)
\sum_Q {\cal{T}}_{pg} (Q) \equiv -\Delta_{pg}^2 G_0(-K)$, which defines
a pseudogap parameter, $\Delta_{pg}$.  This last approximation
is valid only because (through Eq.~(1)) ${\cal{T}}_{pg}$ diverges as
$Q \rightarrow 0$.
In this way, $\Sigma_{pg} (K)$ has a BCS-like form, as does the total
self energy $\Sigma (K) = - \Delta^2 G_0(-K)$, where $\Delta ^2 =
\Delta_{sc}^2 + \Delta_{pg}^2$.  Thus, in the present approach, the
energy gap for single electron excitations reflects the presence of both
finite center-of-mass momentum pairs as well as the condensate. While
the structure of the gap equation will be seen to be formally identical
to that in BCS theory, the vanishing of the excitation gap, $\Delta$,
takes place at a higher temperature than that at which the order
parameter, $ \Delta_{sc}$, vanishes. The latter defines $T_c$.

If we now expand ${\cal {T}}_{pg}^{-1}({\bf q}, \Omega) \approx a_1
\Omega^2 + a_0\Omega +\tau_0 - B q^2 + i\Gamma_{\bf q}^\prime $, we see
that the chemical potential of the pairs $\mu_{pair}$ is proportional to
$\tau_0$, and via Eq.~(1), precisely zero at and below $T_c$.  This
provides an interpretation 
%of the superconducting transition 
along the lines of ideal Bose gas condensation. (Here, also, at small
${\bf q }$, $\Gamma_{\bf q }^\prime \rightarrow 0$).
As $g$ increases, the term $a_0\Omega$ in ${\cal T}_{pg}^{-1}$ becomes
progressively dominant with respect to $a_1 \Omega^2$.  For the
physically relevant regime of moderate $g$, we have found, after
detailed numerical calculations, that $a_1$ may be safely neglected.  At
weak coupling, there is no loss of generality in approximating ${\cal
  T}_{pg}$ in this more particle-hole asymmetric way,
since its contribution  is negligible. In this way, we can write 
\begin{equation} 
{\cal T}_{pg}^{-1}({\mathbf{q}}, \Omega)= a_0 (\Omega - \Omega_{\mathbf{q}} +
\mu_{pair}+ i\Gamma_{\mathbf{q}}) \:. 
\end{equation}

As a consequence we have
\begin{equation}
  \label{eq:gap3}
   \Delta_{pg}^2 = -\sum_Q {\cal{T}}_{pg}(Q) =
\frac{1}{a_0} \sum_{ {\bf q} \ne 0}  b (\Omega_{\bf q}) \:.
\end{equation}
We now rewrite Eq.~(1), along with the fermion number constraint, as
\begin{eqnarray}
  \label{eq:gap1}
  1+g\sum_{\mathbf{k}} \frac{1- 2f(E_{\mathbf{k}})}{2E_{\mathbf{k}}}\,
  \varphi^2_{\mathbf{k}} &=& 0 \:, \\ 
 \label{eq:gap2}
 \sum_{\mathbf{k}}\left[1- \frac{\epsilon_{\mathbf{k}}}{E_{\mathbf{k}}} +
   \frac{2\epsilon_{\mathbf{k}}}{E_{\mathbf{k}}}\,f(E_{\mathbf{k}})\right]
 &=& n \:.
\end{eqnarray} 
Here $f(x)$ and $b(x)$ are the Fermi and Bose functions and
$E_{\mathbf{k}} = \sqrt{ \epsilon_{\mathbf{k}} ^2 + \Delta^2
  \varphi_{\mathbf{k}}^2}$ is the quasiparticle dispersion.  Equations
(3)-(5) are consistent with BCS theory at small $g$, and with the ground
state $\Psi_0$ at all $g$; in both cases the right hand side of Eq.~(3)
is zero.
The simplest physical interpretation of the present decoupling scheme is
that it goes beyond the standard BCS mean field treatment of the single
particles (which also acquire a self energy from the finite {\bf q}
pairs) but it \textit{treats the pairs at a self consistent, mean field
  level}.

The dispersion $\Omega_{\mathbf{q}} = q^2 B / a_0$, as well as the
coefficient $a_0$, are determined by a Taylor expansion of
${\cal{T}}_{pg}^{-1}$ \cite{Chen2}
\begin{eqnarray}
 {\cal T}^{-1}_{pg}({{\mathbf{q}},\Omega}) &=& g^{-1} +
    \sum_{\mathbf{k}} \left[
      \frac{1-f(E_{\mathbf{k}})-f(\epsilon_{{\mathbf{k}}-{\mathbf{q}}})}
      {E_{\mathbf{k}} + \epsilon_{{\mathbf{k}}-{\mathbf{q}}} - \Omega}
      u_{\mathbf{k}}^2 \right.\nonumber\\ 
&&\left. {} -
      \frac{f(E_{\mathbf{k}}) - f(\epsilon_{{\mathbf{k}}-{\mathbf{q}}})}
      {E_{\mathbf{k}} - \epsilon_{{\mathbf{k}}-{\mathbf{q}}} + \Omega}
      v_{\mathbf{k}}^2 \right]
    \varphi_{{\mathbf{k}}-{\mathbf{q}}/2}^2\;.
\end{eqnarray}
For small $g$ and in three dimensions (3D), the poles of
${\cal{T}}_{pg}$ (at $T=0$) occur at $\Omega = \sqrt{3} c q$, where $c$
is the usual phase mode velocity.  At moderate $g$, where the pairons
become increasingly more relevant, and for quasi-2D dispersion
$\epsilon_{\bf k}$, $\Omega_{\mathbf{q}} \approx q_\parallel^2 / 2
M^*_\parallel +q_\perp^2/2M^*_\perp $.  Here we find that the ratio
$M^*_\parallel/M^*_\perp \propto (t_\perp/t_\parallel)^2$, where
$t_\parallel$ and $t_\perp$ are the in- and out-of-plane hopping
integrals, respectively.  Numerical calculations show that the masses,
as well as the residue $a_0$, are roughly $T$ independent constants at
low $T$\cite{Mpair-T}.  In the BEC regime at low density and with
$s$-wave pairing in a 3D continuous model, $M^*$ is $2m_e$ for all
$T\leq T_c$\cite{Chen2}, as found previously\cite{Haussmann}.  The
examples in this paper, which apply to the fermionic regime, correspond
to somewhat smaller $M^*_\parallel$.

It is important to note that in strictly 2D, the logarithmic divergence
on the right hand side of the pseudogap equation (3), (which is
essentially a boson number equation), implies $T_c=0$, as in an ideal
Bose gas.  For large anisotropy, or small $t_\perp$, $T_c \propto -
1/\ln (t_\perp/t_\parallel) $, which vanishes logarithmically
\cite{Chen2}.  Finally, since both $a_0$ and the effective pair mass
(tensor) $M^*$ are constants at low $T$, Eq.~(\ref{eq:gap3}) implies
$\Delta_{pg}^2(T) = \Delta^2(T) - \Delta_{sc}^2(T) \propto T^{3/2}$.
Moreover, because $\Delta$ depends on $T$ only exponentially,
$\Delta_{sc}^2(T) = \Delta^2(0) - A T^{3/2}$ at low $T$, where $A$ is
$T$ independent.

We now calculate physical quantities such as the magnetic penetration depth
($\lambda$) and related superfluid density ($n_s$), the Knight shift ($K_s$)
and NMR relaxation rate ($R_s$) using techniques similar to those used to
study fluctuation effects in normal metal superconductors\cite{Patton}. Here
the usual (lowest order) Maki-Thompson and Aslamazov-Larkin diagrams are
extended to be compatible with $\Sigma$ and $\cal{T}$ 
by applying the generalized Ward identity to incorporate the pairon vertex
correction\cite{Patton,Kosztin2}. Finally, the specific heat $C_v$ can be
computed following a similar analysis as for the paramagnon problem
\cite{Brinkman}.

While in the BCS limit the expressions for $\lambda$, (or $n_s$), $K_s$
and $R_s$ contain only the total gap $\Delta$, here they, in principle,
depend on both the quasiparticle (via $\Delta^2$) and the pairon (via
$\Delta_{pg}^2$) contributions\cite{Kosztin,Kosztin2}. This
decomposition leads to a form of ``three fluid'' model (including the
condensate, fermionic quasi-particles and bosonic pairons).  In the same
way, $C_v$ can also be decomposed into a sum of two contributions
corresponding to an ideal Bose gas of pairons, with dispersion
$\Omega_{\mathbf{q}}$, and an ideal Fermi gas of quasiparticles, with
dispersion $E_{\mathbf{k}}$.  The pairon contributions to $n_s$
 enter as follows\cite{Chen,Kosztin}: the general expression is
identical to its BCS counterpart, but with the overall multiplicative
factor of $\Delta^2$ replaced by $\Delta_{sc}^2 = \Delta^2 -
\Delta_{pg}^2$.  By contrast, for spin-singlet pairing, there is no
explicit pairon contribution to $K_s$ and $R_s$, and the corresponding
expressions reflect the generalized excitation gap $\Delta$, as might
have been expected physically. The single most important conclusion of
this analysis is that \textit{the presence of low lying pair excitations
  will introduce new low temperature power law dependences with ideal
  Bose gas character into physical quantities}.  Below, we explore these
power laws in the context of highly anisotropic 3D, i.e., quasi-2D
systems.

Figures 1(a) and 1(b) present a comparison between an $s$-wave short
$\xi$ pseudogap (PG) superconductor and an $s$- and $d$-wave BCS system.
It should be noted that the short $\xi$ superconductors are still far
from the BEC limit.  For the parameters illustrated by the figures,
$\mu$ deviates from $E_F$ by roughly 3\%. Here, and throughout this
paper we take $t_\perp/t_\parallel = 0.01$.  The main body of Fig.~1(a)
indicates that the Knight shift (and NMR relaxation rate, not shown) at
$T_c$ are substantially reduced relative to their high $T$ asymptotes,
i.e., $K_n$, as is illustrated by the solid line (for the PG $s$-wave).
Because pairon effects are not explicit, the low $T$ behavior is
exponentially activated as for the BCS $s$-wave case, but here the ratio
$\Delta (0) / T_c$ is significantly enhanced over the BCS value.
Overall, the behavior of $R_s$ will yield rather similar plots; however,
the $s$-wave BCS limit exhibits the well-known Hebel-Slichter peak,
which is absent below $T_c$ for the other two cases.

\begin{figure}
\centerline{\includegraphics[clip, width=3.5in]{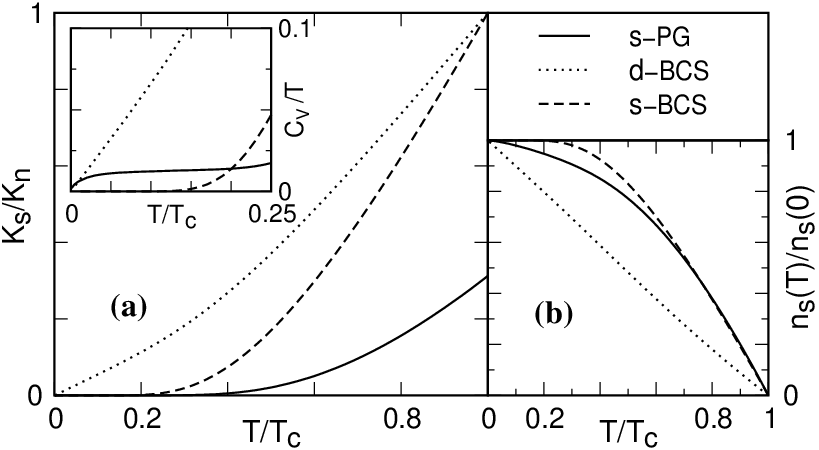}}
\medskip
\caption{Temperature dependence of (a) Knight shift, $K_s$, specific heat 
  $C_v/T$ (inset), and (b) superfluid density $n_s$ in conventional $s$- and
  $d$-wave (BCS) and short $\xi$ (PG) $s$-wave cases, calculated at ($n$,
  $-g/4t_\parallel$)= : $s$-BCS: (0.5, 0.5); $s$-PG: (0.5, 0.7); $d$-BCS:
  (0.8, 0.225); $d$-PG: (0.92, 0.56). }
\end{figure}

In the inset to Fig.~1(a) we plot the behavior of the low $T$ specific heat
``coefficient'', $\gamma(T)\equiv C_v / T$, for the same parameters as above.
In the short $\xi$, quasi-2D case, slightly above $T=0$, $ \gamma(T)$ will
appear to be a constant $\gamma(T) = \gamma ^*$, although it vanishes
strictly at $T=0$ as $T^{1/2}$.  This intrinsic $\gamma^*$ effect, which may
have been seen in both organic and cuprate (layered)
superconductors\cite{Phillips}, has, in the past, been related to extrinsic
effects.
Figure 1(b) plots the normalized superfluid density $n_s$, or $\lambda^{-2}$,
%the inverse square of the penetration depth $\lambda$, 
vs $T/T_c$, which for the PG ($s$-wave) case
exhibits a $T^{3/2}$ dependence.  Here $\lambda$, (unlike $C_v$), is not
particularly sensitive to the mass anisotropy ratio, and the boson power law
dependence is more 3D.

In order to address $d$-wave effects in short $\xi$ superconductors, we
turn to the cuprates. Note, the dimensionless coupling is
$g/t_\parallel$.  To be consistent with the observed metal-insulator
transition at half filling ($x=0$), we introduce a hole concentration
$x$ dependent renormalization of the in-plane hopping integral
$t_\parallel(x) = t_0 x$ deriving from Coulomb correlations, and
presume, in the absence of any more detailed information about the
pairing mechanism, that $g$ is $x$ independent.  Our quasi-2D band
structure is taken from the literature\cite{t0}; the one free parameter
$-g/4t_0$ is chosen ($=0.045$) to optimize agreement with the energy
scales in the cuprate phase diagram.  This calculated phase diagram
\cite{Chen}, deduced from Eqs.~(3)-(5), can be shown to yield reasonable
agreement with experimental data.  Two important points should be
stressed: (i) the chemical potential $\mu / E_F $ differs from unity by
at most a few percent over the entire range of $x$; (ii) While the band
mass increases with underdoping due to Coulomb effects, the
thermodynamically measured mass, obtained from, say, the Knight shift
$K_s(T_c)$ (or specific heat $C_v/T$ at $T_c^-$) \textit{decreases} with
underdoping, as a consequence of the opening of the pseudogap. [See
Fig.~2(c)].

\begin{figure}
\centerline{\includegraphics[clip, width=3.5in]{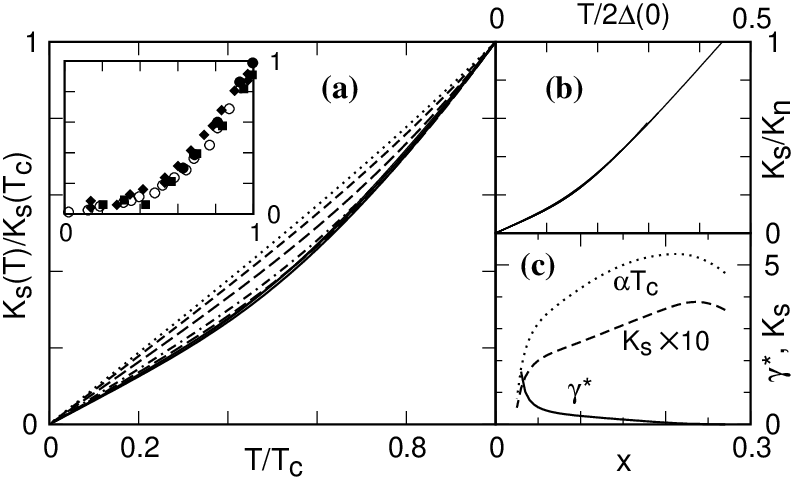}}
\medskip
\caption{(a)-(b) Scaling behavior of the $T$ dependence of Knight
  shift, $K_s$, for the cuprates with respect to doping $x$, and (c) doping
  dependence of $\gamma (T)= \gamma^*+\alpha T$ and $K_s$ at $T_c^-$.
  In (a) [and (b)], $x$ varies from 0.05 to 0.2 from top to bottom. For
  comparison, shown in the inset are experimental data from
  Ref.~\protect\onlinecite{K_data} on under- ({\tiny $\blacksquare$}),
  optimally ({\large $\circ$}, {\large $\bullet$}), and overdoped
  ({\scriptsize $\blacklozenge$}) $\mathrm{Bi_2Sr_2CaCu_2O_{8+\delta}}$
  single crystals. As per the experimental convention for $\varphi_{\bf k}$,
  we use $2\Delta(0)$ in (b).}
\end{figure}

Figures 2(a) and 2(b) illustrate the predicted behavior of the Knight
shift for the cuprates as a function of $x$ and $T$.  Because it depends
only on ($d$-wave nodal) quasiparticle excitations, $K_s$ exhibits a
scaling with $T/\Delta(0)$ which is illustrated in Fig.~2(b) via plots
of $K_s$ (normalized to its high $T$ asymptote $K_n$) for the entire
range of $x$, and for temperatures below each respective $T_c(x)$.  An
alternate scaling form is shown in Fig.~2(a) where we plot $K_s$
normalized at $T_c$ as a function of $T/T_c$ for various $x$ (with $x$
increasing from top to bottom).  The near-collapse of the different
$x$-dependent curves is similar to that found in the experimental
data\cite{K_data} shown in the inset.  The normalization factor
$K_s(T_c)$ for this figure, (which varies as the band mass multiplied by
$T_c / \Delta(0)$), is plotted as a function of $x$ in Fig.~2(c). Also
plotted here are our specific heat ($C_v / T$) predictions for the
pairon contribution to $\gamma ^*$ as compared with the usual $d$-wave
quasiparticle term\cite{Phillips} $ \alpha T_c$ as a function of $x$.
The pairon term becomes increasingly more important with underdoping.

\begin{figure}
\centerline{\includegraphics[width=3.2in, clip]{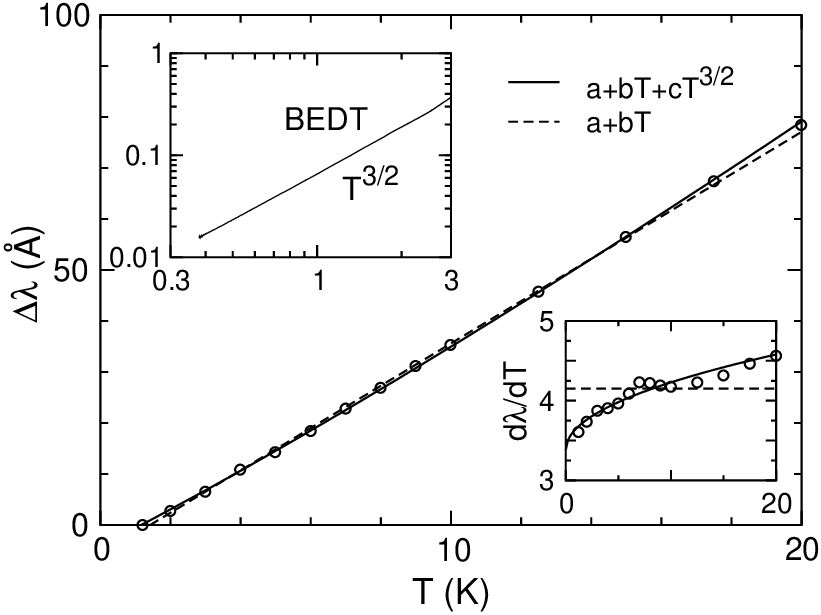}}
\medskip
\caption{Comparison of penetration depth data\protect\cite{Hardy},
  $\Delta\lambda$, along $a$-axis, in nominally pure YBCO$_{6.95}$
  single crystal, with different theoretical fits corresponding to BCS
  $d$-wave (dashed curve) and to BCS-BEC (solid curve) predictions. The
  corresponding derivatives are plotted in the lower inset.  In the upper
  inset are experimental data ($\Delta\lambda$ vs $T$) for the organic
  superconductor BEDT from Ref.~\protect\onlinecite{Russ}.}
\end{figure} 

In Fig.~3 we present $a$-axis penetration depth data, $\Delta\lambda(T) $,
in nominally clean optimally doped YBCO single crystal, from
Ref.~\onlinecite{Hardy}, along with $d$-wave fits to our BCS-BEC theory and
to a straight line associated with a BCS superconductor.  Because these two
fitted curves are essentially indistinguishable, in the lower inset we plot
the slopes $d\lambda / dT $ where the difference between the two sets of
curves is more apparent.  Here it is shown that the low temperature downturn
of the derivative, seen to a greater or lesser extent in all
$\Delta\lambda(T)$ measurements, fits our predicted $T^{1/2} + const.$
dependence rather well.  This downturn has been frequently associated with
impurity effects, which yield a linear in $T$ slope for $\Delta \lambda $ at
very low $T$, and in this case, provide a poorer fit.  While these cuprate
experiments were performed on a nearly optimal sample, the same analysis of
an underdoped material yielded similarly good agreement, but with a
$T^{3/2}$ coefficient about a factor of two larger.  Future more precise and
systematic low $T$ experiments on additional underdoped samples are needed.
Plotted in the upper inset are data\cite{Russ} on the organic
superconductor $\kappa$-(ET)$_2$Cu[N(CN)$_2$]Br (BEDT, $T_c\approx 11$K)
which fit a pure $T^{3/2}$ power law over a wide temperature regime; in
contrast to the cuprates, there is no leading order linear term.  At
present, there seems to be no other explanation (besides the pairon
mechanism presented here) for this unusual power law at the lowest 
temperatures.

In summary, within a BCS-BEC crossover theory, (based on the Leggett
ground state), we find that new low $T$ power laws associated with a
quasi-ideal gas of bosonic pair excitations appear in the thermodynamic
and transport properties, which may be generally relevant to short $\xi$
superconductors.

This work was supported by grants from the NSF under awards No.~DMR~91-20000
(through STCS) and No.~DMR-9808595 (through MRSEC). We thank A. Carrington,
W.~N. Hardy, R.~W. Giannetta, G.~F.  Mazenko, N.~E.  Phillips,
P.~B.  Wiegmann, and especially A.~J. Leggett for useful conversations.


\begin{thebibliography}{10}
\bibitem{Mpair-T}
At low $T$, we find that the leading order corrections are: $\Delta M_{pair}
  \propto T^2$ for $s$-wave and $\Delta M_{pair} \propto T$ for $d$-wave;
  $\Delta a_0 \propto T^2$ in both cases; These correction terms are all very
  small.
\end{thebibliography}
\end{document}